\documentclass[twocolumn,showkeys,prl,nofootinbib]{revtex4-1}

\usepackage{amsfonts,amsmath,amssymb}
\usepackage{graphicx,graphics}
\usepackage{epsfig}
\usepackage{natbib}

\def\mnras{MNRAS}
\def\apj{ApJ}
\def\apjl{ApJL}
\def\apjs{ApJS}

\def\aap{A\&A}

\def\araa{Ann. Rev. A\&A}

\begin{document}

\title{On the Neutron Star-Black Hole Binaries Produced by Binary-driven-Hypernovae}

\author{Chris L.~Fryer,$^{1}$ F.~G.~Oliveira,$^{2,3,4}$ J.~A.~Rueda,$^{2,3,4,5}$, R.~Ruffini$^{2,3,4,5}$}
\affiliation{$^1$CCS-2, Los Alamos National Laboratory, Los Alamos, NM 87545}
\affiliation{$^2$Dipartimento di Fisica and ICRA, Sapienza Universit\`a di Roma, P.le Aldo Moro 5, I--00185 Rome, Italy}
\affiliation{$^3$ICRANet, Piazza della Repubblica 10, I--65122 Pescara, Italy}
\affiliation{$^4$Universit\'e de Nice - Sophia Antipolis Cedex 2, Grand Ch\^ateau Parc Valrose, Nice, France}
\affiliation{$^5$ICRANet-Rio, Centro Brasileiro de Pesquisas F\'isicas, Rua Dr. Xavier Sigaud 150, Rio de Janeiro, RJ, 22290--180, Brazil}
\date{\today}

\begin{abstract}
Binary-driven-hypernovae (BdHNe) within the induced gravitational
collapse (IGC) paradigm have been introduced to explain energetic
($E_{\rm iso}\gtrsim 10^{52}$~erg), long gamma-ray bursts (GRBs)
associated with type Ic supernovae (SNe). The progenitor is a tight
binary composed of a carbon-oxygen (CO) core and a neutron star (NS)
companion, a subclass of the newly proposed ``ultra-stripped''
binaries.  The CO-NS short-period orbit causes the NS to accrete
appriciable matter from the SN ejecta when the CO core collapses,
ultimately causing it to collapse to a black hole (BH) and producing a
GRB.  These tight binaries evolve through the SN explosion very
differently than compact binaries studied in population synthesis
calculations.  First, the hypercritical accretion onto the NS
companion alters both the mass and momentum of the binary.  Second,
because the explosion timescale is on par with the orbital period, the
mass ejection can not be assumed to be instantaneous.  This
dramatically affects the post-SN fate of the binary.  Finally, the bow
shock created as the accreting NS plows through the SN ejecta
transfers angular momentum, braking the orbit.  These systems remain
bound even if a large fraction of the binary mass is lost in the
explosion (well above the canonical 50\% limit), and even large kicks
are unlikely to unbind the system. Indeed, BdHNe produce a new family
of NS-BH binaries unaccounted for in current population synthesis
analyses and, although they may be rare, the fact that nearly 100\%
remain bound implies they may play an important role in the compact
merger rate, important for gravitational waves (GWs) that, in turn, can 
produce a new class of ultrashort GRBs.
\end{abstract}

\keywords{Type Ic Supernovae --- Hypercritical Accretion --- Induced Gravitational Collapse --- Gamma Ray Bursts -- Gravitational Waves}

\maketitle

\section{Introduction}\label{sec:1}

Binary massive star systems evolve into a broad set of compact
binaries from X-ray binaries consisting of stars accreting onto either
BH or NS companions to the more exotic binary
compact objects such NS-BH and NS-NS binaries. The formation scenarios of these
compact binaries typically argue that, after the first SN
explosion, the compact remnant enters a common envelope phase with its
companion, tightening the orbit. If the system remains bound after
the companion star collapses, a NS-BH or NS-NS binary is formed. A
range of scenarios have been invoked \cite{1999ApJ...526..152F,2012ApJ...759...52D,2014LRR....17....3P},
including exotic scenarios where both components expand off the main
sequence concurrently, causing a single common envelope around two
helium cores \cite{1995ApJ...440..270B}.

Recently, two independent communities have argued for a ``new''
  evolutionary scenario forming these compact binaries where, after
  the collapse of the primary star to a NS, the system undergoes a
  series of mass transfer phases, ejecting both the hydrogen and
  helium shells of the secondary to produce a binary composed of a
  massive CO core and a NS.  When the CO core collapses and produces a
  SN explosion, a compact binary system is formed.  In the X-ray
  binary/SN community, these systems are called ``ultra-stripped''
  binaries.  In the past few years, such systems have been invoked to
  both explain the population of NS-NS binaries as well as a growing
  set of low-luminosity and/or rapid decay-rate SNe
  \cite{2013ApJ...778L..23T,2015MNRAS.451.2123T}.  Low-mass ejecta can
  match the observational features of these SNe and ultra-stripped
  binaries without hydrogen and helium layers in their pre-SN
  progenitor produce small cores with such low-mass explosions.  The
  rate of these systems are predicted to be 0.1--1\% of the total SN
  rate \cite{2013ApJ...778L..23T}.  These binaries are extremely
  tight, and most of the systems studied have orbital periods lying
  between $3000$ and $300,000$~s.  Proponents of the ultra-stripped
  systems argue that this scenario dominates the formation of NS-NS
  binaries and that there are virtually no systems that are formed
  where the CO core collapses directly to a BH.

The IGC scenario for GRBs
  \cite{2001ApJ...555L.117R,2008mgm..conf..368R,2012ApJ...758L...7R}
  introduced a subset of extremely short-period CO-NS binaries where
  the ejecta from the exploding CO star accretes onto its NS
  companion, causing the NS, in some cases, to collapse to a BH.  If
  ultra-stripped binaries dominate the formation of NS-NS binaries,
  this scenario would dominate the formation of NS-BH binaries.  This
  collapse to a BH releases energy to drive the GRB emission
  \cite{2012ApJ...758L...7R,2014ApJ...793L..36F}.  The CO core is a
  requirement to allow the tight orbits needed to produce sufficient
  accretion to cause the NS collapse, but it also provides a natural
  explanation for the fact that these GRBs are always associated with
  type Ic SNe.  The recently introduced ultra-stripped binaries are a
  welcome support for the IGC scenario from the point of view of
  stellar evolution, with the only caveat that IGC progenitors are a
  small subset of the ultra-stripped binaries where the initial
  orbital separation and CO core mass are aligned to produce binaries
  with orbital periods lying in the 100--1000\,s range.  This requires
  fine-tuning both of the CO star mass and the binary orbit. From an
  astrophysical point of view the IGC scenario is uniquely
  characterized by the formation of the BH during the accretion
  process of the SN ejecta onto the companion NS and the associated
  GRB emission. Since the rate of the high-luminosity GRBs (BdHNe)
  explained through the IGC scenario is $(1.1$--$1.3)\times
  10^{-2}$~Gpc$^{-3}$~y$^{-1}$ \cite{MuccinoRates}, and 0.1--1\% of
  the SN Ibc population could be ultra-stripped binaries
  \cite{2013ApJ...778L..23T}, only 0.005--0.07\% of the latter are
  needed to explain the BdHNe population (assuming a SN Ibc rate of
  $2\times 10^4$~Gpc$^{-3}$~y$^{-1}$ \cite{2007ApJ...657L..73G}).

Studies of ultra-stripped binaries have expanded our
  understanding of stellar radii, confirming these results: CO cores
  with masses below 2~$M_\odot$ have radii of $1$--$4\times 10^9$~cm
  \cite{2015arXiv150608827S}, in agreement with the assumptions used
  in IGC studies \cite{2014ApJ...793L..36F}.  Even if some helium
  remains on the stripped core, it will be ejected if it expands to
  interact with its compact-object companion.  These radii are
  sufficiently small to produce the tight orbits required to produce
  the rapid accretion of the ejecta onto the NS companion and the
  formation of the BH.

In typical systems, most of the binaries become unbound during the
SN explosion because of the ejected mass and momentum imparted
(kick) on the newly formed compact object in the explosion of the
massive star.  Under the instantaneous explosion assumption, if half of the binary
system's mass is lost in the SN explosion, the system is disrupted,
forming two single compact objects.  Although SN kicks may allow
some systems to remain bound, in general, these kicks unbind even more
systems.  In general, it is believed that the fraction of massive
binaries that can produce double compact object binaries is low:
$\sim$0.001--1\%~\cite{1999ApJ...526..152F,2012ApJ...759...52D,2014LRR....17....3P}.

For ultra-stripped binaries, the fate is very different.  In these
systems, the mass ejected is extremely low and, if the SN kick
is low, these systems remain bound \cite{2013ApJ...778L..23T,2015MNRAS.451.2123T}.  In the
tighter binaries leading to IGC progenitors, the assumption of instantaneous
mass ejection is no longer valid.  We demonstrate in this work
that, removing this assumption, even with a strong SN kick
nearly all of these systems will remain bound.  In this case, even
though IGC progenitors are rare, the compact binaries produced by these
progenitors may dominate the total NS-BH binaries in the Universe, and
lead to a new previously unaccounted family of GRBs.

We shall describe below the differences between these systems and
typical massive star binaries, modeling these orbits through the
SN explosion. We then calculate the evolution of these NS-BH
binaries via GWs emission up to the merger point, and
assess their detectability. We conclude with a discussion of the
additional observational predictions of these NS-BH binaries,
introducing a new class of short GRBs, with specific observational
signatures, here referred to as ultrashort GRBs.

\section{Post-Explosion Orbits}\label{sec:orbits}

The mass ejected during the SN alters the binary orbit,
causing it to become wider and more eccentric. Assuming that
the mass is ejected instantaneously, the post-explosion semi-major 
axis is $a/a_0=(M_0 - \Delta M)/(M_0 - 2 a_0 \Delta M/r)$,
%
%
where $a_0$ and $a$ are the initial and final semi-major axes respectively,
$M_0$ is the total initial mass of the binary system, $\Delta M$ is
the change of mass (equal to the amount of mass ejected in the
SN), and $r$ is the orbital separation at the time of
explosion \cite{1983ApJ...267..322H}.  For circular orbits, like the ones expected from our
systems after going through a common envelope evolution, we
find that the system is unbound if it loses half of its mass.  But,
for these close binaries, a number of additional effects can
alter the fate of the binary.

The time it takes for the ejecta to flow past a companion in a SN is
roughly 10--1000~s.  These explosions follow a so-called
  homologous velocity profile where the velocity is proportional to
  the position.  Although the shock front is moving above
  10,000\,km\,s$^{-1}$, the denser,lower-velocity ejecta can be moving
  at below 1000\,km\,s$^{-1}$.  Our estimates are based on simulated
  supernova explosions \cite{2014ApJ...793L..36F}. The broad range of
  times arises because the SN ejecta velocities varies from
  100--10,000~km~s$^{-1}$.  The accretion peaks as the slow-moving
  (inner) ejecta flows past the NS companion. Note that the initial SN
  explosion in this case is not a hypernova.  The observed
  ``hypernova'' is actually produced when the GRB from the BH collapse
  sweeps up this SN (and circumstellar) material \cite{2015ApJ...798...10R}.  For normal
binaries, this time is a very small fraction of the orbital period and
the ``instantaneous'' assumption is perfectly valid.  However, in the
close binary systems considered here, the orbital period ranges from
only 100--1000~s, and the mass loss from the SN explosion can no
longer be assumed to be instantaneous.

\begin{figure}
\centering
\includegraphics[width=0.9\hsize,clip]{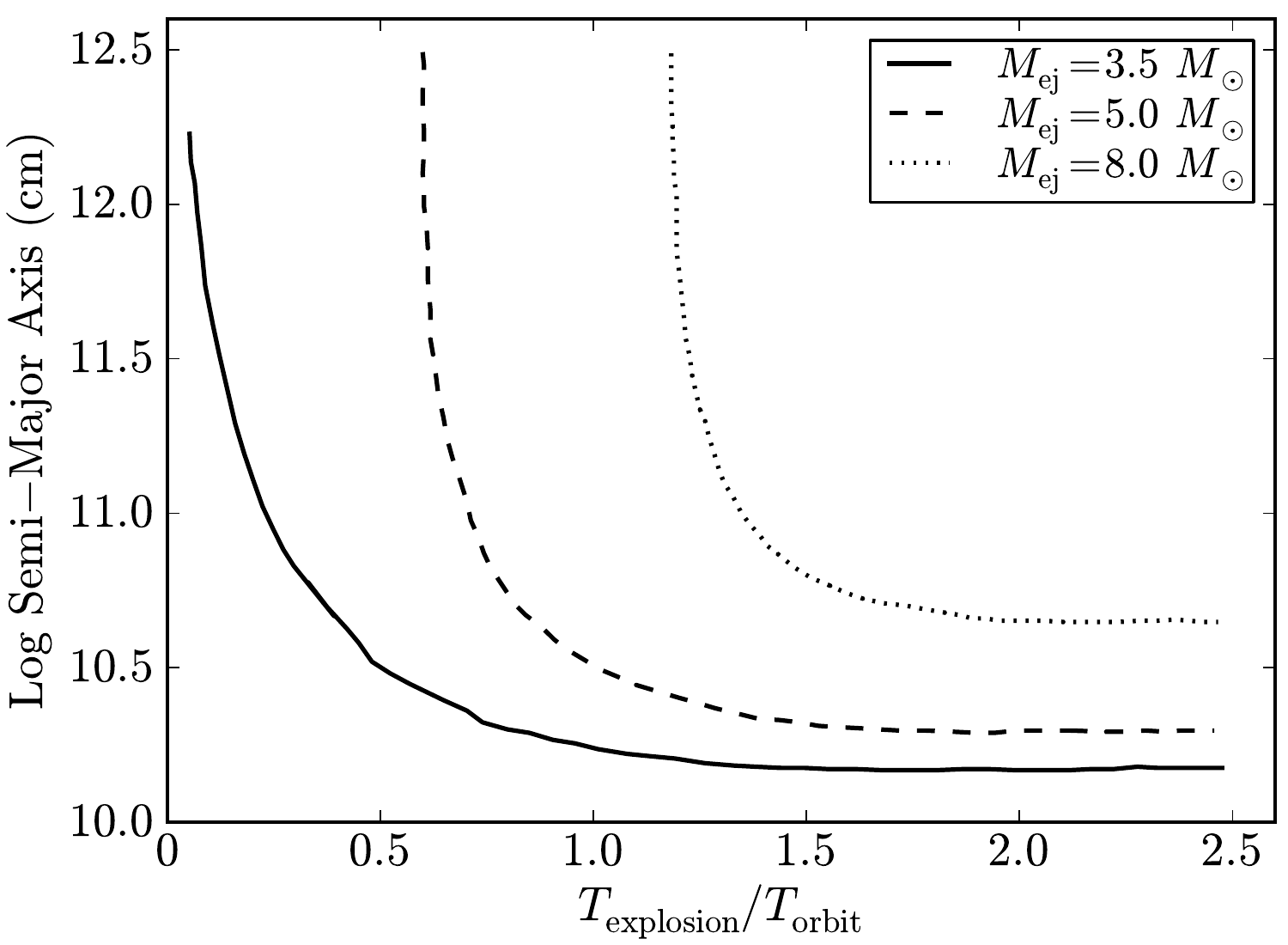}
\caption{Semi-major axis versus explosion time for 3 different mass
  ejecta scenarios: 3.5\,$M_\odot$ (solid), 5.0\,$M_\odot$ (dotted),
  8.0\,$M_\odot$ (dashed). The CO core collapse to form a
  1.5\,$M_\odot$ NS (its initial mass is the ejecta mass
  plus the NS mass), and the companion NS has a
  mass of 2.0\,$M_\odot$. If the explosion were instantaneous, all of
  our systems with ejecta masses above 3.5\,$M_\odot$ would be unbound.
  For explosion times above 1.2 times the orbital time, not only are
  the systems bound, but the final orbital semi-major axis is less
  than 10 times the initial separation.}
\label{fig:orb1}
\end{figure}

This has already been pointed out in \cite{2015ApJ...812..100B} where it was shown that in BdHNe the accretion process is fast and massive enough to produce the BH formation in a time-interval as short as the orbital period. We here deepen this analysis to study the effect of the SN explosion in such a scenario with a specific example, for which we have
produced an orbit code using a simple staggered leapfrog
integration (see \cite{1991ppcs.book.....B} for details of this integration method). We have tested both stability (by
modeling many orbits) and convergence (decreasing the time step by 2
orders of magnitude confirming identical results). We also reproduce
the results of the instantaneous
limit. From figure~\ref{fig:orb1}, as the ejecta timescale becomes
just a fraction of the orbital timescale, the fate of the
post-explosion binary can be radically altered. For these models, we
assumed very close binaries with an initial orbital separation of
$7\times10^9\,{\rm cm}$ in circular orbits (such close binaries are
only formed through a common envelope phase which circularizes the
orbit). With CO core radii of $1$--$4 \times 10^9$~cm \cite{2015arXiv150608827S},
such a separation is small, but achievable.  We assume the binary
consists of a CO core and a 2.0~$M_\odot$ NS companion.
When the CO core collapses, it forms a 1.5~$M_\odot$ NS,
ejecting the rest of the core.  We then vary the ejecta mass and time
required for most of the ejected matter to move out of the binary.  Note that
even if 70\% of the mass is lost from the system (the 8~$M_\odot$
ejecta case), the system remains bound as long as the explosion time
is just above the orbital time ($T_{\rm orbit}=180$~s) with semi-major
axes of less than $10^{11}$~cm.

\begin{figure}
\centering
\includegraphics[width=0.9\hsize,clip]{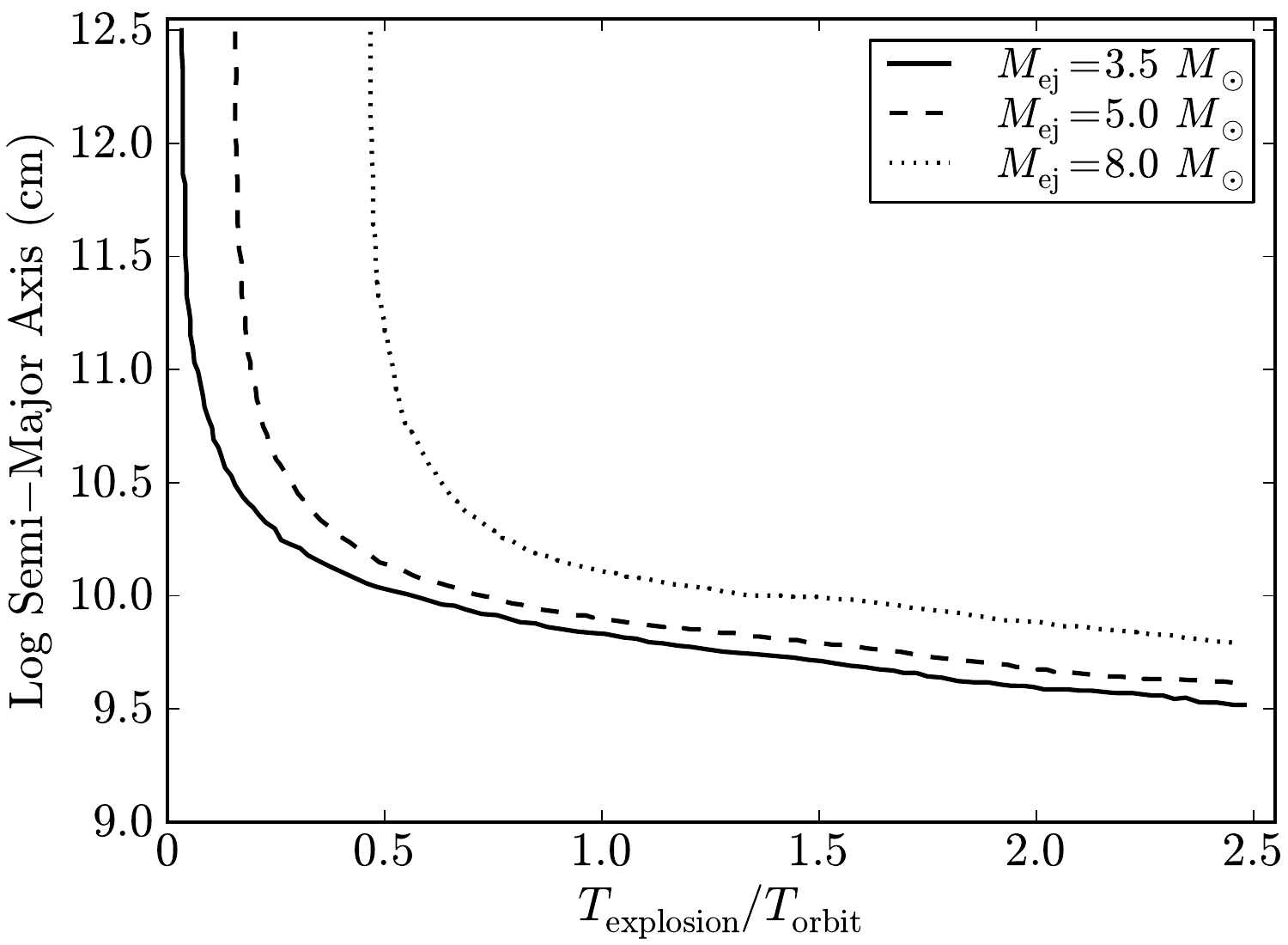}
\caption{Semi-major axis versus explosion time for the same 3 binary 
systems as in figure~\ref{fig:orb1} including mass accretion 
and momentum effects. Including these effects, all systems with 
explosion times above 0.7 times the orbital time are bound and the 
final separations are on par with the initial separations.}
\label{fig:orb2}
\end{figure}

The short orbits (on ejecta timescales) are not the only feature
of these binaries that alters the post-explosion orbit.  The NS companion accretes both matter and momentum from the SN
ejecta, reducing the mass lost from the system with respect to typical binaries with larger orbital separations and much less accretion. In addition, as with common envelope scenarios, the bow 
shock produced by the accreting NS transfers orbital 
energy into the SN ejecta. In figure~\ref{fig:orb2}, we show 
the final orbital separation of our same three binaries, including the 
effects of mass accretion (we assume 0.5\,$M_\odot$ is accreted 
with the momentum of the SN material) and orbit coupling 
(30\% of the orbital velocity is lost per orbit).  With these 
effects, not only do the systems remain bound even for explosion 
times greater than 1/2 the orbital period but, if the explosion time 
is long, the final semi-major axis can be on par with the initial 
orbital separation.

The tight separation of these binaries facilitates tidal locking and
the angular momentum axis of the CO core will be aligned with the
orbital angular momentum.  For many of the kick mechanisms in the
literature, the kick is often aligned with the rotation axis. 
For example, both in neutrino-driven mechanisms \cite{2000ApJ...541.1033F,2006ApJS..163..335F} and asymmetric explosions driven by convection \cite{1994ApJ...432L.119S,2000ApJ...541.1033F,2004ApJ...601..391F} the kick
is aligned with the rotation axis. However, it is still possible to have some misalignment leading to some eccentricity and ``tumbling'' of the system with specific signatures in the light curve following the prompt emission of the GRB.
Hence, we here consider both kicks aligned with the
rotation (and hence orbital) axis as well as random kicks.
If the kick is aligned with the orbital plane, the system can remain bound even 
with kick velocities as high as 1000~km~s$^{-1}$.  However, if the kick is in 
the same direction as the star is moving, the systems can be disrupted if 
the kick is above 500--700~km~s$^{-1}$ if the accretion phase is longer than 
an orbital period.

\begin{figure}
\centering
\includegraphics[width=0.9\hsize,clip]{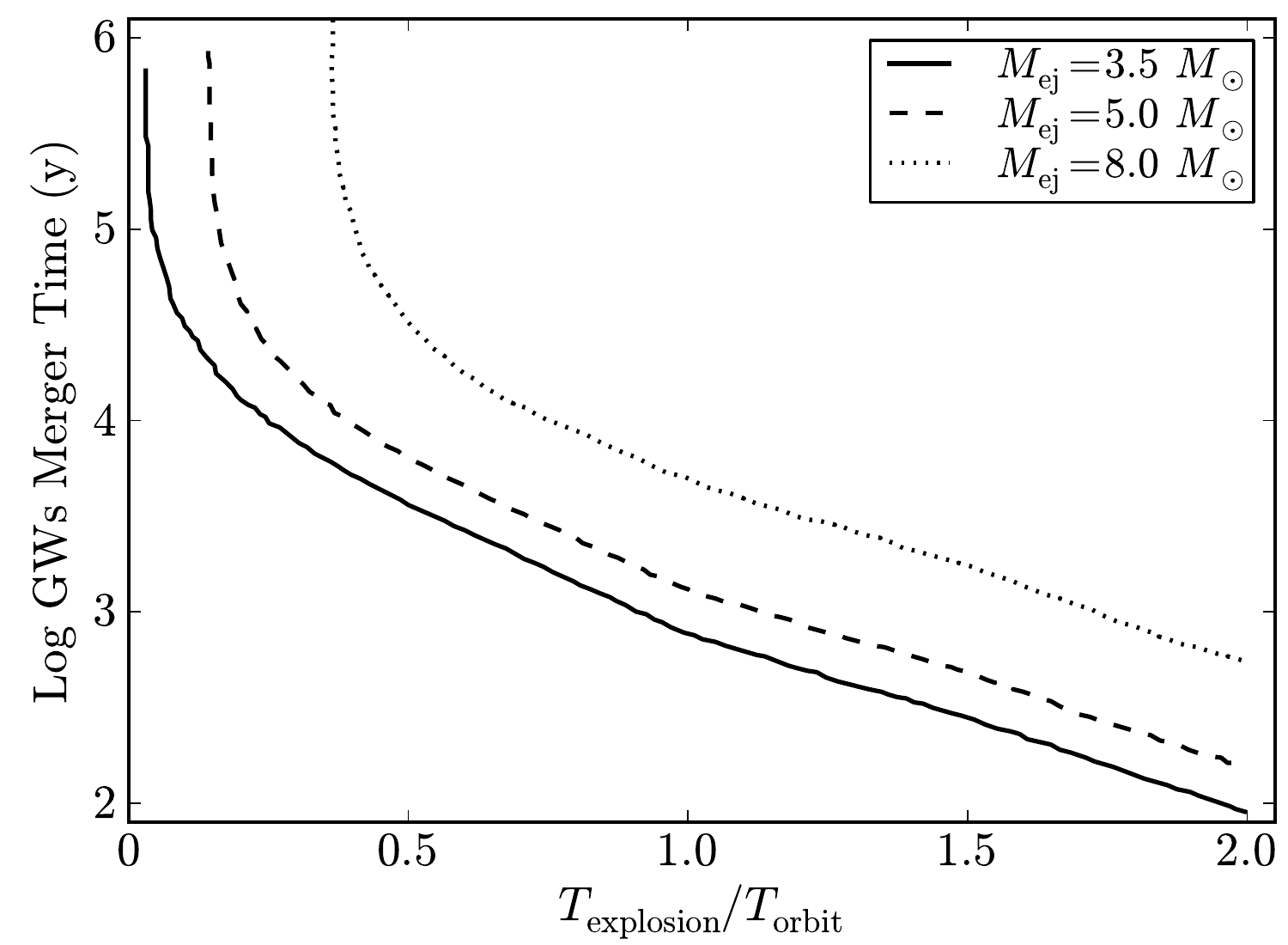}
\caption{Merger time due to GW emission as a function of explosion time for the same 3
  binary as in figure~\ref{fig:orb1} including mass accretion and
  momentum effects.  Beyond a critical explosion time (0.1--0.6~$T_{\rm
    orbit}$ depending on the system), the merger time is less than
  roughly 10,000~y.  For most of our systems, the explosion time is
  above this limit and we expect most of these systems to merge
  quickly.}
\label{fig:merger}
\end{figure}

The tight compact binaries produced in these explosions will emit
GW emission, ultimately causing the system to merge.
For typical massive star binaries, the merger time is many Myr.  For
BdHNe, the merger time is typically 10,000~y, or less (figure~\ref{fig:merger}).


\section{Gravitational Waves from the NS-BH binary}\label{sec:gw}

To better understand the GW signal from these mergers, we study the
evolution of the orbital binding energy $E_b$ up to the merger
following the effective one-body (EOB) formalism
\cite{1999PhRvD..59h4006B,2000PhRvD..62f4015B,2000PhRvD..62h4011D,2001PhRvD..64l4013D}
up to the 4th Post-Newtonian approximation (see
Refs.~\cite{2013PhRvD..87l1501B,2014ApJ...787..150O} and references
therein). We adopt here $M_{\rm NS}=1.5~M_\odot$ and $M_{\rm
  BH}=2.67~$M$_\odot$ \cite{2012NuPhA.883....1B}, the latter
corresponding to the critical mass $M_{\rm crit}$ of a non-rotating NS
obeying the nuclear NL3 equation of state (EOS). Uncertainties
  in the EOS at supranuclear densities lead to a variety of NS
  mass-radius relations and consequently to different values of
  $M_{\rm crit}$, hence of $M_{\rm BH}$. Both rotation
  \cite{2015PhRvD..92b3007C} and different binary parameters may lead
  to different amounts of angular momentum transferred to the NS,
  affecting its mass \cite{2015arXiv150507580B}.

\begin{figure}[!hbtp]
\includegraphics[width=\hsize,clip]{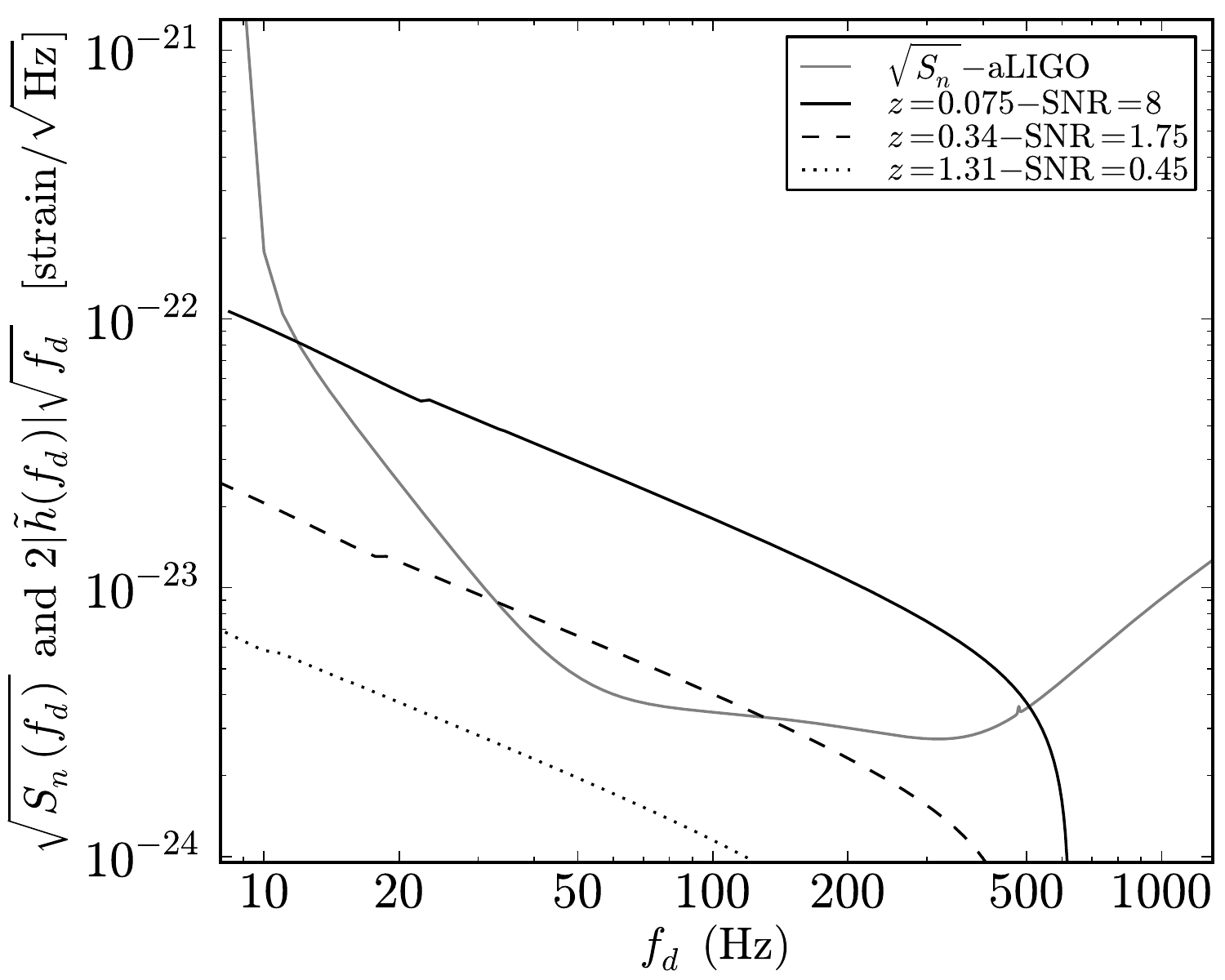}
\caption{ASD of the spiraling-in phase up to the merger, $\sqrt{S_h}=2 |\tilde{h}(f_d)|\sqrt{f_d}=h_c(f_d)/\sqrt{f_d}$, of the NS-BH binaries produced by two BdHNe, GRB 130427A at redshift $z=0.34$ and GRB 061121 at $z=1.31$, compared with the noise ASD of aLIGO, $\sqrt{S_n}(f_d)$. We indicate the estimated SNR for these two sources and show the case of the NS-BH binary which would generate a positive detection with SNR=8. The binary dynamics is simulated via the EOB formalism up to the 4th Post-Newtonian approximation.}
\label{fig:SNR}
\end{figure}

In order to assess the detectability of the GW emission by advanced LIGO (aLIGO), we compute the signal-to-noise ratio (SNR), averaged over all sky locations and binary orientations, $\langle{\rm SNR}\rangle$, generated by the NS-BH spiraling-in binary up to the merger point \cite{2014ApJ...787..150O}.
%
%
Following \cite{2013arXiv1304.0670L}, we adopt as a threshold for the aLIGO detection $\langle{\rm SNR}\rangle=8$ in a single detector, which implies a GW horizon distance for these NS-BH binaries, which have a chirp mass $\mathcal{M}_{\rm ch}=(M_{\rm BH} M_{\rm NS})^{3/5}/(M_{\rm BH} + M_{\rm NS})^{1/5}\approx 1.73~M_\odot$, $d_L\approx 335.4$~Mpc, or $z\approx 0.075$, using the maximum possible sensitive reachable by 2022. No BdHN has been up to now detected with such a low redshift. Figure~\ref{fig:SNR} shows, for two sources shown to be consistent with the BdHN picture (GRB 130427A with $z=0.34$ \cite{2015ApJ...798...10R} and GRB 061121 with $z=1.31$ \cite{2014A&A...569A..39R}), the GW source amplitude spectral density, $\sqrt{S_h}=2 |\tilde{h}(f_d)|\sqrt{f_d}=h_c(f_d)/\sqrt{f_d}$, together with the one-sided ASD of the aLIGO noise, $\sqrt{S_n}(f_d)$. Here $h_c(f_d)$ and $\tilde{h}(f_d)$ are the characteristic strain and the Fourier transform of the signal and $f_d$ the frequency of the GWs at the detector. For these sources, $\langle{\rm SNR}\rangle \approx 1.75$ and 0.45, respectively. For an optimally located and polarized source, the SNR could increase by up to a factor $\approx 2.26$, which implies that SNR=8 could be obtained for a source as far as $d_L\approx 2.26\times 335.4$~Mpc~$\approx 758$~Mpc, or $z\approx 0.160$. Furthermore, the SNR scales as $\mathcal{M}^{5/6}_{\rm ch}$, so it increases e.g. with larger BH masses. For rotating NS with the NL3 EOS, the maximum value of $M_{\rm crit}$ is $\approx 3.4~M_\odot$ \cite{2015PhRvD..92b3007C}, which would increase the SNR only by $\approx 1.1$. For this largest BH mass, the GW horizon becomes $d_L\approx 1.1\times 758$~Mpc~$\approx 834$~Mpc, or $z\approx 0.174$. This largest possible GW horizon implies an upper limit of $\sim 0.03$ detections per year, adopting a BdHN rate of $1.2\times 10^{-2}$~Gpc$^{-3}$~y$^{-1}$ \cite{MuccinoRates}.

%

\section{Conclusion}

The evolutionary scenario for BdHNe requires much tighter binaries
than typically studied in the literature of ultra-stripped binaries and this produces unique
features in the end-fate of these systems. The progenitor of this GRB
engine begins with two massive stars, in contrast to the one based on a massive core collapsing to a BH \cite{2006ARA&A..44..507W}.  A tight
binary is produced after a succession of common envelope phases,
producing a CO core near Roche-Lobe overflow orbiting a NS, a subset of the 
ultra-stripped binaries \cite{2001ApJ...550L.183B,2003MNRAS.344..629D,2013ApJ...778L..23T,2015MNRAS.451.2123T}. Since 0.1--1\% of the total SN Ibc are expected to be ultra-stripped binaries \cite{2013ApJ...778L..23T}, we estimate that only 0.005--0.07\% of the latter
are needed to explain the observed population of BdHNe. The fate of such systems evolves very differently than the
standard picture. The NS can accrete appreciable material in the
SN explosion and this accretion causes it to
collapse to a BH and form a GRB.  However, the tight binary invalidates many
of the assumptions about orbital evolution in the SN.  The
SN explosion does not pass ``instantaneously'' across the NS,
and correcting this assumption alone drastically alters the binary fate.  Including the interaction of the orbit and the ejecta 
further exacerbates these differences, causing these NS-BH 
to be very different than the systems prediction in standard population 
synthesis models.

First and foremost, the fraction of the BdHNe that remain
bound after the SN explosion is nearly 100\% even with large
$\sim 500$--$1000$~km~s$^{-1}$ kicks imparted during the SN 
explosion instead of the $\lesssim$1\% in standard scenarios.  This
means that even if BdHNe are rare, they may dominate the fraction of NS-BH binaries 
in the Universe. In addition, the merger timescales
for these systems are typically $<$10,000\,y, producing a set of
rapidly-merging binaries. In view of such a short lifetime due to GW emission, the current number of such events is likely to be comparable with the 
original rate of long GRBs produced by BdHNe following the IGC paradigm. Because of this rapid merger, 
the systems are unlikely to travel that far from the site of the SN
explosion that formed the GRB.  Even with large kicks, we expect
these binaries to merge within 10\,pc of the BdHNe and we expect the
merger to occur within the radius swept clean by the BdHN, giving a characteristic imprint in the GRB emission.  In view
of the expected paucity of the baryonic contamination around the
merger site, it is expected that the characteristic prompt radiation emission time of the GRB produced by these sources be dominated by the general relativistic timescale of the BH,
$G M/c^3\approx 10^{-4}$--$10^{-5}$~s,
which justifies the attribution of the name of ultrashort GRBs to
this new family of events.


Another observational feature of these binaries is that the BHs from these systems are low
mass: $\sim$3--4\,$M_\odot$, of the order of the critical mass of rotating NSs \cite{2015PhRvD..92b3007C,2015arXiv150507580B}, instead of the
5--10\,$M_\odot$ produced by standard scenarios. However, further accretion of mass and angular momentum from material kept bound into the system after the BdHN process might lead the BH to larger masses and to approach maximal rotation \cite{2015arXiv150507580B}. Although the NS
in this NS-BH binary should be rapidly rotating, producing pulsed
emission, the short timescale between formation and merger means that
it will be difficult to observe such systems through steady pulsed
emission. However, if these systems make up a sizable fraction of the
NS-BH population, they could be detecting by their GW
signal.  Although it is difficult to get the exact component masses
from aLIGO, evidence \cite{2013ApJ...766L..14H}, or the lack
thereof, for binaries with low-mass BHs could support, or
limit the rate of, this scenario.\\

We would like to thank P. Podsiadlowski, T. Tauris and Y. Suwa for
many useful discussions about ultra-stripped binaries. Likewise, we would like to thank D. Arnett and S. O. Kepler for insightful discussions.
J.~A.~R. acknowledges the support by the International Cooperation
Program CAPES-ICRANet financed by CAPES – Brazilian Federal Agency
for Support and Evaluation of Graduate Education within the Ministry
of Education of Brazil. F.~G.~O. acknowledges the support given by the
International Relativistic Astrophysics Erasmus Mundus Joint Doctorate
Program under the Grant 2012–-1710 from EACEA of the European
Commission.


%

\end{document}